\documentclass[12pt]{article}
\usepackage{amsmath,amsfonts,amsthm}

\newcommand{\AS}{{A^\star}}

\newcommand{\DS}{{D^\star}}
\newcommand{\TS}{{T^\star}}
\newcommand{\Complex}{{\mathbb C}}
\newcommand{\Integer}{{\mathbb Z}}

\newcommand{\Span}[1]{\mbox{Span \{}#1\mbox{\}}}
\newcommand{\IdOp}{{\mbox{\bf 1}}}
\renewcommand{\vec}[1]{\underline{#1}}

\title{Matrix difference equations for the supersymmetric Lie algebra $sl(2,1)$ and the `off-shell' Bethe ansatz}

\author{T. Quella\footnote{E-mail: quella@physik.fu-berlin.de}$\,\;$\footnote{New adress since November 1999: Max-Planck-Institut f\"ur
  Gravitationsphysik, Albert-Einstein-Institut, Am M\"uhlenberg 1, D-14476
  Golm, Germany, E-mail: quella@aei-potsdam.mpg.de}\\Institut f\"ur
  theoretische Physik, Freie
  Universit\"at Berlin, \\Arnimallee 14, D-14195 Berlin, Germany}

\date{25. November 1999}

\begin{document}

\maketitle

\vspace{-8cm}\hspace{10cm}SFB-288-427, hep-th/9911213\vspace{7cm}

\begin{abstract}
  Based on the rational $R$-matrix of the supersymmetric~$sl(2,1)$ matrix difference
  equations are solved by means of a generalization of the nested algebraic Bethe ansatz.
  These solutions are shown to be of highest-weight with respect to the
  underlying graded Lie algebra structure.
\end{abstract}

% -------------------------------------------------------------------------
% -------------------------------------------------------------------------
% -------------------------------------------------------------------------
\section{Introduction}

  The supersymmetric t-J model is often considered a candidate for describing
  high $T_c$
  superconductivity~\cite{Korepin:1994,Foerster:1993fp,Foerster:1993uk}. The
  underlying symmetry is described by the supersymmetric (graded) Lie algebra
  $sl(2,1)$. Integrable models with supersymmetry have been discussed also
  in~\cite{Bassi:1999ua, Gohmann:1999av, Pfannmuller:1996vp, Ramos:1996my,
  Maassarani:1995ac, Hinrichsen:1992nj}.
  This article extends the results in~\cite{KarowskiSU(N),Zapletal:1998qw} on
  matrix difference equations and a generalized version of the algebraic Bethe
  ansatz for ordinary or quantum groups to this supersymmetric Lie algebra.
  We recall that matrix difference equations play an important role in
  mathematical physics, see
  e.g.~\cite{Babujian:1998uw,Smirnov:1992,TarasovVarchenko,Frenkel:1992gx}. In
  particular in the context of quantum integrable field theories they provide
  solutions of the formfactor equations, which can be used to calculate
  correlation functions~\cite{Berg:1979sw}. This type of matrix
  difference equations can also be considered as a discrete version~\cite{Reshetikhin92} of a 
  Knizhnik-Zamolodchikov system~\cite{Knizhnik:1984nr}.

  The conventional algebraic Bethe ansatz is used to solve the eigenvalue
  problem of a hamiltonian in a way closely related to the underlying symmetry of the considered
  model~(see e.g.~\cite{Faddeev:1981}). One constructs the
  eigenvectors as highest
  weight vectors of the corresponding irreducible representations either of the
  ordinary Lie algebra or the $q$-deformed analogue, the quantum group. By this
  construction one encounters `unwanted' terms. The eigenvalue equation is
  fulfilled if all these `unwanted' terms vanish which leads to the so called
  Bethe ansatz equations.

  The `off-shell' Bethe ansatz~\cite{BabujianI, BabujianII, Reshetikhin92,
  KarowskiSU(N), Zapletal:1998qw}
  is used to solve matrix differential or difference equations. The solution
  is represented as an integral or a sum over some lattice (integral of
  Jackson type). The `unwanted'
  terms arising in this case do not vanish due to the
  Bethe ansatz equations but they sum up to zero under the integral or sum.
  This modification of the Bethe ansatz has originally been introduced to
  solve Knizhnik-Zamolodchikov
  equations~\cite{BabujianI}. It has also been applied to
  the quantization of dimensionally reduced gravity~\cite{Korotkin:1996vi} in
  this connection.

  Let $f^{1\cdots n}(\vec{x}):\Complex^n\rightarrow V^{1\cdots
  n}=\bigotimes_{j=1}^{n}\Complex^3$ be a vector valued function having the following
  symmetry property
\begin{eqnarray}
  \label{eq:SSym}
  f^{\cdots ij\cdots}(\cdots,x_i,x_j,\cdots)
  =R_{ji}(x_j-x_i)f^{\cdots ji\cdots}(\cdots,x_j,x_i,\cdots),
\end{eqnarray}
  where~$R$ is the $sl(2,1)$ $R$-matrix (see below).
  We consider the set of matrix difference equations
\begin{eqnarray}
  \label{eq:QPeriod}
  f^{1\cdots n}(x_1,\cdots,x_i+\xi,\cdots,x_n)
  =Q_{1\cdots n}(\vec{x}|i)f^{1\cdots n}(\vec{x})\hspace{1cm}(i=1,\cdots,n)
\end{eqnarray}
  with an arbitrary shift-parameter~$\xi$ and some sort of generalized tranfer
  matrix~$Q_{1\cdots n}(\vec{x}|i)$ invariant under~$sl(2,1)$.
  Functions satisfying~\eqref{eq:SSym} and~\eqref{eq:QPeriod} will be called
  $R$-symmetric and $Q$-periodic respectively.

% -------------------------------------------------------------------------
% -------------------------------------------------------------------------
% -------------------------------------------------------------------------
\section{Matrix difference equation and generalized nested Bethe ansatz}
  \label{ch:MaDiffEq}

  Let $V^{1\cdots n}=V_1\otimes\cdots\otimes V_n$ be the tensor product of~$n$
  isomorphic vector spaces $V_i=\Span{|1\rangle,|2\rangle,|3\rangle}\cong\Complex^3$.
  The states~$|1\rangle$ and~$|2\rangle$ are supposed to be bosonic
  while~$|3\rangle$ is fermionic~\cite{Foerster:1993uk}.
  For later convenience we also define the reduced vector spaces
  $\tilde{V}_i=\Span{|2\rangle,|3\rangle}\cong \Complex^2$
  and $\tilde{V}^{1\cdots m}=\tilde{V}_1\otimes\cdots\otimes \tilde{V}_m$.
  Vectors in~$V^{1\cdots n}$ will be denoted by $f^{1\cdots n}\in V^{1\cdots
  n}$. Analogously vectors in the reduced spaces are in addition marked with
  a tilde: $\tilde{f}^{1\cdots m}\in \tilde{V}^{1\cdots m}$.
  Matrices acting in~$V^{1\cdots n}$ are denoted with index subscripts
  $Q_{1\cdots n}: V^{1\cdots n}\rightarrow V^{1\cdots n}$.

  As usual the $R$-matrix will depend on a spectral parameter~$\theta$. This
  matrix $R_{ij}(\theta):V_i\otimes V_j\rightarrow V_j\otimes V_i$ is of the
  form~\cite{Bracken:1990qy}
\begin{eqnarray}
  %\label{eq:SMatrix}
  R_{ij}(\theta)=b(\theta)\Sigma_{ij}+c(\theta)P_{ij}
\end{eqnarray}
  where $P_{ij}: |\alpha\rangle\otimes|\beta\rangle\mapsto|\alpha\rangle\otimes|\beta\rangle$ is the permutation operator
  and
\begin{eqnarray*}
  \Sigma_{ij}: |\alpha\rangle\otimes|\beta\rangle\mapsto\sigma_{\alpha\beta}|\beta\rangle\otimes|\alpha\rangle
      =\left\{\begin{array}{rl}
         -|\beta\rangle\otimes|\alpha\rangle &, |\alpha\rangle=|\beta\rangle=|3\rangle\\
         |\beta\rangle\otimes|\alpha\rangle &, \mbox{ else}.
       \end{array}\right.
\end{eqnarray*}
  The statistics factor~$\sigma_{\alpha\beta}=\pm1$ takes
  the fermionic character of the state $|3\rangle$ into account. It has the value~$-1$ if and
  only if both states are fermionic, i.e.~$\alpha=\beta=3$. The
  functions~$b(\theta)$ and~$c(\theta)$ have the form
\begin{eqnarray*}
  b(\theta)=\frac{\theta}{\theta+K}\hspace{4cm}c(\theta)=\frac{K}{\theta+K}
\end{eqnarray*}
  with an arbitrary constant~$K$.
  For later use we define the function $w(\theta)=-b(\theta)+c(\theta)$.
  It is easy to check that $R(\theta)$ is unitary and satisfies the Yang-Baxter
  equation:
\begin{eqnarray}
  \label{eq:UnitaritySYBE}
  R_{ab}(\theta)R_{ba}(-\theta)=\IdOp\quad\mbox{ and }\quad
  R_{12}(\theta_{12})R_{13}(\theta_{13})R_{23}(\theta_{23})
  =R_{23}(\theta_{23})R_{13}(\theta_{13})R_{12}(\theta_{12}),
\end{eqnarray}
  where $\theta_{ij}=\theta_i-\theta_j$.

  Next we introduce different kinds of monodromy matrices which prove
  to be useful in the following.
  The monodromy matrix
\begin{eqnarray*}
  T_{1\cdots n,a}(\vec{x}|u)=R_{1a}(x_1-u)\cdots R_{na}(x_n-u).
\end{eqnarray*}
  is an operator $V^{1\cdots n}\otimes V_a \rightarrow V_a\otimes V^{1\cdots
  n}$. The vector spaces $V^{1\cdots n}$ and $V_a$ are called quantum and
  auxiliary space respectively. As usual we will consider this operator as a matrix
\begin{eqnarray*}
  T_{1\cdots n,a}=\left(\begin{array}{ccc}
                     A & B_2 & B_3 \\
                     C^2 & D_2^2 & D_3^2 \\
                     C^3 & D_2^3 & D_3^3
                  \end{array}\right)
\end{eqnarray*}
  over the auxiliary space with operators in the quantum space as entries.
  As a consequence of~\eqref{eq:UnitaritySYBE} the monodromy
  matrix fulfills the Yang-Baxter algebra relation
\begin{eqnarray}
  %\label{eq:YBE}
  R_{ab}(u-v)
  T_{1\cdots n,b}(\vec{x}|v)T_{1\cdots n,a}(\vec{x}|u)
  =T_{1\cdots n,a}(\vec{x}|u)T_{1\cdots n,b}(\vec{x}|v)
   R_{ab}(u-v).
\end{eqnarray}

  Following~\cite{KarowskiSU(N)} we also introduce another set of modified monodromy
  matrices for~$i=1,\cdots,n$ given as
\begin{eqnarray*}
  T_{1\cdots n,a}^Q(\vec{x}|i)=R_{1a}(x_1-x_i)\cdots R_{i-1a}(x_{i-1}-x_i) P_{ia}
  R_{i+1a}(x_{i+1}-x_i^\prime)\cdots R_{na}(x_n-x_i^\prime),
\end{eqnarray*}
  where $\vec{x}\,^\prime=\vec{x}+\xi\vec{e}_i$. In the same way as above they
  should be considered as matrices in the auxiliary space.
  This new type of monodromy matrix satisfies the two mixed Yang-Baxter relations
\begin{eqnarray*}
  T_{1\cdots n,a}^Q(\vec{x}|i)T_{1\cdots n,b}(\vec{x}|u)R_{ab}(x_i^\prime-u)
  &=&R_{ab}(x_i-u)T_{1\cdots n,b}(\vec{x\,}^\prime|u)
     T_{1\cdots n,a}^Q(\vec{x}|i)\\
  T_{1\cdots n,a}(\vec{x\,}^\prime|u)T_{1\cdots n,a}^Q(\vec{x}|i)
     R_{ab}(u-x_i^\prime)
  &=&R_{ab}(u-x_i)T_{1\cdots n,b}^Q(\vec{x}|i)T_{1\cdots n,a}(\vec{x}|u).
\end{eqnarray*}
  For~$i=n$ the modified monodromy matrix is the same as the ordinary one.

  We want to encode the fermionic nature of the state~$|3\rangle$ in such a
  way that $sl(2,1)$ appears naturally.
  To do so we define an additional monodromy matrix
\begin{eqnarray}
  %\label{eq:FMod}
  [\TS_{1\cdots n,a}(\vec{x}|u)]_{\alpha,\{\mu\}}^{\beta,\{\nu\}}
  =\sigma_{\alpha\beta}\sigma_{\beta\nu_1}\cdots\sigma_{\beta\nu_n}
   [T_{1\cdots n,a}(\vec{x}|u)]_{\alpha,\{\mu\}}^{\beta,\{\nu\}},
\end{eqnarray}
  where the quantum space indices are collected in the
  notation~$\{\nu\}=\nu_1,\cdots,\nu_n$.
  This definition is easily extended to a modified version as before.
  The shift operator is defined by
\begin{eqnarray}
  \label{eq:Shiftmatrix}
  Q_{1\cdots n}(\vec{x}|i)=tr_a \TS_{1\cdots n,a}^Q(\vec{x}|i)=A_{1\cdots
  n,a}^Q(\vec{x}|i)+\sum_{\alpha=2,3} \left[\DS_{1\cdots n,a}^Q(\vec{x}|i)\right]_\alpha^\alpha.
\end{eqnarray}
  which is obviously closely related to usual transfer matrices.
  For all operators just defined there also exists a counterpart in the reduced
  spaces denoted by a tilde.

  Using the Yang-Baxter relations given above we derive in a straightforward
  way the commutation relations
\begin{eqnarray}
  \label{eq:VTRone}
  B_i(\vec{x}|u_2)B_j(\vec{x}|u_1)
  &=&B_{j^\prime}(\vec{x}|u_1)B_{i^\prime}(\vec{x}|u_2)
      R_{ji}^{i^\prime j^\prime}(u_1-u_2)\\
%  \label{eq:VTRtwo}
%  B_j(\vec{x\,}^\prime|u)B_k^Q(\vec{x}|i)
%  &=&B_{k^\prime}^Q(\vec{x}|i)B_{j^\prime}(\vec{x}|u)
%      R_{kj}^{j^\prime k^\prime}(x_i-u)\\
  \label{eq:VTRthree}
  A(\vec{x}|u_2)B_i(\vec{x}|u_1)
  &=&\frac{1}{b(u_2-u_1)}B_i(\vec{x}|u_1)A(\vec{x}|u_2)
     -\frac{c(u_2-u_1)}{b(u_2-u_1)}B_i(\vec{x}|u_2)A(\vec{x}|u_1)\\
  \label{eq:VTRfour}
  A^Q(\vec{x}|i)B_j(\vec{x}|u)
  &=&\frac{1}{b(x_i^\prime-u)}B_j(\vec{x\,}^\prime|u)A^Q(\vec{x}|i)
     -\frac{c(x_i^\prime-u)}{b(x_i^\prime-u)}B_j^Q(\vec{x}|i)A(\vec{x}|u)\\
  \label{eq:VTRfive}
  \DS_j^i(\vec{x}|u_2)B_k(\vec{x}|u_1)
  &=&\frac{\sigma_{ik}}{b(u_1-u_2)}B_{k^\prime}(\vec{x}|u_1)\DS_{j^\prime}^i(\vec{x}|u_2)
      R_{kj}^{j^\prime k^\prime}(u_1-u_2)\\
   &&\qquad\qquad\qquad\qquad-\sigma_{ik}\frac{c(u_1-u_2)}{b(u_1-u_2)}B_j(\vec{x}|u_2)\DS_k^i(\vec{x}|u_1)\nonumber\\
  \label{eq:VTRsix}
  \DS_k^{Qj}(\vec{x}|i)B_l(\vec{x}|u)
  &=&\sigma_{jl}\frac{1}{b(u-x_i)}B_{l^\prime}(\vec{x\,}^\prime|u)
      \DS_{k^\prime}^{Qj}(\vec{x}|i)R_{lk}^{k^\prime l^\prime}(u-x_i^\prime)\\
   &&\qquad\qquad\qquad\qquad-\sigma_{jl}\frac{c(u-x_i)}{b(u-x_i)}B_k^Q(\vec{x}|i)\DS_l^j(\vec{x}|u)\nonumber
\end{eqnarray}
  The first terms on the right hand side of each of these equations are called
  `wanted' the others `unwanted'. These relations are slightly different than
  those appearing in the $SU(N)$-case~\cite{KarowskiSU(N)} due to the
  statistics factors~$\sigma$ in the last two equations.

  To solve the system of~\eqref{eq:SSym} and the matrix difference
  equations~\eqref{eq:QPeriod} we use the nested so called `off shell' Bethe
  ansatz~\cite{BabujianI,BabujianII} with two levels. The first level is quite analogous
  to the constructions in~\cite{KarowskiSU(N),Zapletal:1998qw}. Due to the
  fermionic statistics of state~$|3\rangle$ which ensures supersymmetry the second level
  is different. This problem is solved in the present paper. We write the vector valued
  function~$f^{1\cdots n}:\Complex^n\rightarrow V^{1\cdots n}$ as a sum of
  first level Bethe ansatz vectors
\begin{eqnarray}
  \label{eq:BA.1}
  f^{1\cdots n}(\vec{x})
  =\sum_{\vec{u}} B_{\beta_m}(\vec{x}|u_m)\cdots B_{\beta_1}(\vec{x}|u_1)
   \Omega^{1\cdots n}[g^{1\cdots m}(\vec{x}|\vec{u})]^{\beta_1\cdots \beta_m}.
\end{eqnarray}
  The sum is extended over $\vec{u}\in\vec{u}_0-\xi\Integer^m\subset\Complex^m$
  (`integral of Jackson type', $\vec{u}_0\in\Complex^m$ arbitrary).
  The reference state $\Omega^{1\cdots n}$ is given by $\Omega^{1\cdots n}=|1\rangle^{\otimes n}$
  and the auxiliary function
  $g^{1\cdots m}:\Complex^n\times\Complex^m\rightarrow \tilde{V}^{1\cdots m}$
  is defined by
  $g^{1\cdots m}(\vec{x}|\vec{u})
  =\eta(\vec{x}|\vec{u})\tilde{f}^{1\cdots m}(\vec{u})$
  with $\eta:\Complex^n\times\Complex^m\rightarrow\Complex$
\begin{eqnarray*}
  \eta(\vec{x}|\vec{u})
  =\prod_{i=1}^n\prod_{j=1}^m\psi(x_i-u_j)\prod_{1\leq i<j\leq m}\tau(u_i-u_j),
\end{eqnarray*}
  where the scalar functions $\psi:\Complex\rightarrow\Complex$ and
  $\tau:\Complex\rightarrow\Complex$ satisfy
\begin{eqnarray}
  \label{eq:fktglone}
  b(x)\psi(x)=\psi(x-\xi)\hspace{3cm}
  \frac{\tau(x)}{b(x)}=\frac{\tau(x-\xi)}{b(\xi-x)}.
\end{eqnarray}
  Possible solutions are
\begin{eqnarray}
  \label{eq:solutions}
  \psi(x)=\frac{\Gamma(1+\frac{K}{\xi}+\frac{x}{\xi})}{\Gamma(1+\frac{x}{\xi})}
  \qquad\mbox{ and }\qquad
  \tau(x)=x\frac{\Gamma(\frac{x}{\xi}-\frac{K}{\xi})}{\Gamma(1+\frac{x}{\xi}+\frac{K}{\xi})}.
\end{eqnarray}
  They may be multiplied by an arbitrary function which is periodic in~$\xi$.

  We prove that $f^{1\cdots n}(\vec{x})$ is $R$-symmetric and $Q$-periodic if
  $\tilde{f}^{1\cdots m}(\vec{u})$ is $\tilde{R}$-symmetric and $\tilde{Q}$-periodic.
  To compute the action of the shift operator~$Q$ on our Bethe ansatz
  function~$f^{1\cdots n}(\vec{x})$ we start from~\eqref{eq:Shiftmatrix} and commute the
  operators~$A^Q$ und~$\DS^Q$ through all the $B$-operators to the right where
  they act on the reference states according to
  $A^Q(\vec{x}|m)\Omega^{1\cdots n}=\Omega^{1\cdots n}$ and
  $\left[\DS^Q(\vec{x}|m)\right]_{\alpha}^{\alpha^\prime}\Omega^{1\cdots n}=0$. 
  If~$\tilde{f}^{1\cdots m}(\vec{u})$ is $\tilde{R}$-symmetric one obtains the
  representations ($\vec{x\,}^\prime=\vec{x}+\xi\vec{e}_n$)
\begin{multline*}
  \shoveleft{\AS^Q(\vec{x}|n)f^{1\cdots n}(\vec{x})
  =f^{1\cdots n}(\vec{x\,}^\prime)
   +\sum_{\vec{u}}\sum_{i=1}^m
    \Lambda_A^{(i)}(\vec{x}|\vec{u})B_{\beta_i}^Q(\vec{x}|n)}\hspace{5.5cm}\\
  \qquad\times B_{\beta_m}(\vec{x}|u_m)
     \cdots\widehat{B_{\beta_i}(\vec{x}|u_i)}\cdots B_{\beta_1}(\vec{x}|u_1)
    \Omega^{1\cdots n}
    \eta(\vec{x}|\vec{u})\left[\tilde{f}^{1\cdots
  mi}(u_1,\cdots,u_m,u_i)\right]^{\beta_1\cdots\beta_m\beta_i},\\{}
  \shoveleft{[\DS^Q(\vec{x}|n)]_{\alpha}^{\alpha}f^{1\cdots n}(\vec{x})
  =\sum_{\vec{u}}\sum_{i=1}^m\Lambda_D^{(i)}(\vec{x}|\vec{u})B_{\beta_i}^Q(\vec{x}|n)B_{\beta_m}(\vec{x}|u_m)
     \cdots\widehat{B_{\beta_i}(\vec{x}|u_i)}\cdots B_{\beta_1}(\vec{x}|u_1)}\\
   \times\Omega^{1\cdots n}\eta(\vec{x}|\vec{u})\left[\tilde{Q}(u_1,\cdots,u_m,u_i|i)\tilde{f}^{1\cdots mi}(u_1,\cdots,u_m,u_i)\right]^{\beta_1\cdots\beta_m\beta_i}.
\end{multline*}
  The hat denotes a factor which is omitted and $\tilde{Q}(u_1,\cdots,u_m,u_i|i)$
  is an analogue to the shift operator~\eqref{eq:Shiftmatrix} in the dimensionally
  reduced space $\tilde{V}^{1\cdots mi}$. The `wanted' contributions already
  ensure the validity of~\eqref{eq:QPeriod}, so
  `unwanted' ones have to sum up to zero.
  The representation can be obtained
  as follows: The expression in front of the sum is a consequence of the `wanted'
  parts of the commutation relations~\eqref{eq:VTRone}-\eqref{eq:VTRsix}.
  To determine the functions $\Lambda_A^{(i)}(\vec{x}|\vec{u})$ and
  $\Lambda_D^{(i)}(\vec{x}|\vec{u})$ one has to perform the following steps:
  First move $B_{\beta_i}(\vec{x}|u_i)$ to the front of the $B$-operators according
  to~\eqref{eq:VTRone} and use the $\tilde{R}$-symmetry of $\tilde{f}(\vec{u})$ to
  absorb them. Then consider the `unwanted' contributions of~\eqref{eq:VTRfour}
  and~\eqref{eq:VTRsix} respectively. Now commute the resulting
  operators~$A(\vec{x}|u_i)$ and~$\DS(\vec{x}|u_i)$ to the right and only take
  the `wanted' contributions into account. This gives a product of
  $R$-matrices and statistics factors in the case of~$\DS$. The action on the reference state
  is given by $A(\vec{x}|u_i)\Omega^{1\cdots n}=\Omega^{1\cdots n}$ and 
  $[\DS(\vec{x}|u_i)]_{\alpha}^{\alpha^\prime}\Omega^{1\cdots n}=
  \delta_{\alpha}^{\alpha^\prime}\sigma_{\alpha\alpha^\prime}\prod_{j=1}^{n}b(x_j-u_i)
  \Omega^{1\cdots n}$ respectively. By an `ice rule' for fermions
%use of conservation of fermions
  one can regroup the statistics factors. Together with the $R$-matrices this
  gives the reduced shift operator~$\tilde{Q}(u_1,\cdots,u_m,u_i|i)$.
  Finally one obtains
\begin{eqnarray*}
  \Lambda_A^{(i)}(\vec{x}|\vec{u})
  &=&-\frac{c(x_n^\prime-u_i)}{b(x_n^\prime-u_i)}\prod_{l\neq i}\frac{1}{b(u_i-u_l)},\\
  \Lambda_D^{(i)}(\vec{x}|\vec{u})
  &=&-\frac{c(u_i-x_n)}{b(u_i-x_n)}\prod_{l\neq i}\frac{1}{b(u_l-u_i)}\prod_{l=1}^nb(x_l-u_i).
\end{eqnarray*}

  As already mentioned we have to show that the contributions of the sums cancel. 
  Following the arguments of~\cite{KarowskiSU(N)}, i.e. using~\eqref{eq:fktglone} and the relation
  $\frac{c(x)}{b(x)}=-\frac{c(-x)}{b(-x)}$, one can indeed
  show that these `unwanted' contributions vanish after the summation,
  if~$\tilde{f}^{1\cdots m}(\vec{u})$ is $\tilde{Q}$-periodic. The
  symmetry of~$\eta(\vec{x}|\vec{u})$ in the arguments $x_1,\cdots,x_n$
  combined with % the equation
  $R_{ij}(\theta)\Omega^{1\cdots n}=\Omega^{1\cdots n}$
  implies the $R$-symmetry of~$f^{1\cdots n}(\vec{x})$.

  The next step consists in the construction of a function~$\tilde{f}^{1\cdots
  m}(\vec{u})$ which is $\tilde{R}$-symmetric and $\tilde{Q}$-periodic.
  As above we write
\begin{eqnarray}
  \label{eq:BA.2}
  \tilde{f}^{1\cdots m}(\vec{u})
  =\sum_{\vec{v}} \tilde{B}(\vec{u}|v_k)\cdots \tilde{B}(\vec{u}|v_1)
   \tilde{\Omega}^{1\cdots m}\tilde{g}(\vec{u}|\vec{v}).
\end{eqnarray}
  The sum is extended over $\vec{v}\in\vec{v}_0-\xi\Integer^k\subset\Complex^k$ 
  ($\vec{v}_0\in\Complex^k$ arbitrary). Here the reference state is given by
  $\tilde{\Omega}^{1\cdots m}=|2\rangle^{\otimes m}$
  and the auxiliary function
  $\tilde{g}:\Complex^m\times\Complex^k\rightarrow\Complex$ reads
\begin{eqnarray*}
  \tilde{g}(\vec{u}|\vec{v})
  =\prod_{i=1}^m\prod_{j=1}^k\psi(u_i-v_j)
   \prod_{1\leq i<j\leq k}\tilde{\tau}(v_i-v_j),
\end{eqnarray*}
  where $\psi:\Complex\rightarrow\Complex$
  and $\tilde{\tau}:\Complex\rightarrow\Complex$ satisfy
\begin{eqnarray}
  \label{eq:fktgltwo}
  b(x)\psi(x)=\psi(x-\xi)\hspace{3cm}
  \frac{\tilde{\tau}(x)}{b(-x)}=\frac{\tilde{\tau}(x-\xi)}{b(\xi-x)}.
\end{eqnarray}
  Possible solutions of~\eqref{eq:fktgltwo} are given by~\eqref{eq:solutions} and
  $\tilde{\tau}(x)=x/(x-K)$. Again both functions may be multiplied by an
  arbitrary function periodic in~$\xi$. Note that the supersymmetry has modified
  the last equation compared to~\eqref{eq:fktglone}.

  The Yang-Baxter relations imply the commutation relations
\begin{eqnarray*}
  %\label{eq:VTRthree}
  \tilde{B}(\vec{u}|v_2)\tilde{B}(\vec{u}|v_1)
  &=&w(v_1-v_2)\tilde{B}(\vec{u}|v_1)\tilde{B}(\vec{u}|v_2)\\
  %\label{eq:VTRfour}
%  \tilde{B}(\vec{u\,}^\prime|v)\tilde{B}^Q(\vec{u}|i)
%  &=&w(u_i^\prime-v)\tilde{B}^Q(\vec{u}|i)\tilde{B}(\vec{u}|v)\\
  %\label{eq:VTRseven}
  \tilde{A}(\vec{u}|v_2)\tilde{B}(\vec{u}|v_1)
  &=&\frac{1}{b(v_2-v_1)}\tilde{B}(\vec{u}|v_1)\tilde{A}(\vec{u}|v_2)
     -\frac{c(v_2-v_1)}{b(v_2-v_1)}\tilde{B}(\vec{u}|v_2)\tilde{A}(\vec{u}|v_1)\\
  %\label{eq:VTReight}
  \tilde{A}^Q(\vec{u}|i)\tilde{B}(\vec{u}|v)
  &=&\frac{1}{b(u_i^\prime-v)}
      \tilde{B}(\vec{u\,}^\prime|v)\tilde{A}^Q(\vec{u}|i)
     -\frac{c(u_i^\prime-v)}{b(u_i^\prime-v)}
      \tilde{B}^Q(\vec{u}|i)\tilde{A}(\vec{u}|v)\\
  %\label{eq:VTReleven}
  \tilde{\DS}(\vec{u}|v_2)\tilde{B}(\vec{u}|v_1)
  &=&-\frac{w(v_1-v_2)}{b(v_1-v_2)}\tilde{B}(\vec{u}|v_1)\tilde{\DS}(\vec{u}|v_2)%\\
%   &&\qquad\qquad\qquad\qquad
+\frac{c(v_1-v_2)}{b(v_1-v_2)}\tilde{B}(\vec{u}|v_2)\tilde{\DS}(\vec{u}|v_1)\\%\nonumber\\
  %\label{eq:VTRtwelve}
  \tilde{\DS}^Q(\vec{u}|i)\tilde{B}(\vec{u}|v)
  &=&-\frac{w(v-u_i^\prime)}{b(v-u_i)}\tilde{B}(\vec{u\,}^\prime|v)
      \tilde{\DS}^Q(\vec{u}|i)%\\
%   &&\qquad\qquad\qquad\qquad
+\frac{c(v-u_i)}{b(v-u_i)}\tilde{B}^Q(\vec{u}|i)\tilde{\DS}(\vec{u}|v).%\nonumber
\end{eqnarray*}
  Due to supersymmetry these relations are structurally different from those
  for the ordinary group case~\cite{KarowskiSU(N)}.
  As a consequence the function~$\tilde{\tau}$ has to satisfy a slightly modified
  functional equation~\eqref{eq:fktgltwo} compared to~$\tau$ in~\eqref{eq:fktglone}.

  Next we act with the shift operator~$\tilde{Q}^{1\cdots m}(\vec{u}|i)=\tilde{A}^Q(\vec{u}|i)+\tilde{\DS}^Q(\vec{u}|i)$ on the Bethe
  ansatz vector~$\tilde{f}^{1\cdots m}(\vec{u})$ and repeat the arguments
  given above. The equations~\eqref{eq:QPeriod} are equivalent for all
  $i=1,\cdots,m$, so we will restrict ourselves to~$i=m$.
  Using the relations $\tilde{A}^Q(\vec{u}|m)\tilde{\Omega}^{1\cdots m}=\tilde{\Omega}^{1\cdots m}$
  and $\tilde{\DS}^Q(\vec{u}|m)\tilde{\Omega}^{1\cdots m}=0$ we get the
  representations ($\vec{u\,}^\prime=\vec{u}+\xi\vec{e}_m$)
\begin{eqnarray}
  %\label{eq:durchtauschone}
  \tilde{\AS}^Q(\vec{u}|m)\tilde{f}^{1\cdots m}(\vec{u})
  &=&\tilde{f}^{1\cdots m}(\vec{u\,}^\prime)\nonumber\\
   &&\hspace{-2.5cm}+\sum_{\vec{v}}\sum_{i=1}^k \tilde{\Lambda}_A^{(i)}(\vec{u}|\vec{v})\tilde{B}^Q(\vec{u}|m)\tilde{B}(\vec{u}|v_k)
     \cdots\widehat{\tilde{B}(\vec{u}|v_i)}\cdots\tilde{B}(\vec{u}|v_1)\tilde{\Omega}\tilde{g}(\vec{u}|\vec{v})\\
  %\label{eq:durchtauschtwo}
  \tilde{\DS}^Q(\vec{u}|m)\tilde{f}^{1\cdots m}(\vec{u})\nonumber\\
  &&\hspace{-2.5cm}=\sum_{\vec{v}}\sum_{i=1}^k \tilde{\Lambda}_D^{(i)}(\vec{u}|\vec{v})\tilde{B}^Q(\vec{u}|m)\tilde{B}(\vec{u}|v_k)
     \cdots\widehat{\tilde{B}(\vec{u}|v_i)}\cdots\tilde{B}(\vec{u}|v_1)\tilde{\Omega}\tilde{g}(\vec{u}|\vec{v}).
\end{eqnarray}
  By similar arguments as before one can show that the
  functions $\tilde{\Lambda}_A^{(i)}(\vec{u}|\vec{v})$ and
  $\tilde{\Lambda}_D^{(i)}(\vec{u}|\vec{v})$ are given by
\begin{eqnarray*}
  \tilde{\Lambda}_A^{(i)}(\vec{u}|\vec{v})
  &=&-\frac{c(u_m^\prime-v_i)}{b(u_m^\prime-v_i)}\prod_{l<i}\frac{1}{b(v_i-v_l)}\prod_{l>i}\frac{-1}{b(v_l-v_i)}\\
  \tilde{\Lambda}_D^{(i)}(\vec{u}|\vec{v})
  &=&\frac{c(v_i-u_m)}{b(v_i-u_m)}\prod_{l<i}\frac{1}{b(v_i-v_l)}\prod_{l>i}
     \frac{-1}{b(v_l-v_i)}\prod_{l=1}^m b(u_l-v_i).
\end{eqnarray*}
  We made use of the fact that $w(\theta)w(-\theta)=1$ and
  $\frac{w(\theta)}{b(\theta)}=\frac{-1}{b(-\theta)}$.

  Again the `wanted' contributions already guarantee the validity
  of~\eqref{eq:QPeriod} while a straightforward calculation
  using~\eqref{eq:fktgltwo} and~$\frac{c(x)}{b(x)}=-\frac{c(-x)}{b(-x)}$ shows
  that the `unwanted' contributions sum up to zero.
  The $\tilde{R}$-symmetry is implied by the symmetry of
  $\tilde{g}(\vec{u}|\vec{v})$ in the variables~$u_1,\cdots,u_m$ and the
  property
  $\tilde{R}_{ij}(\theta)\tilde{\Omega}^{1\cdots m}=\tilde{\Omega}^{1\cdots m}$.

  Finally we have proved that~$f^{1\cdots n}$ given by the Bethe
  ansatz~\eqref{eq:BA.1} solves the combined system of  
  $R$-symmetry~\eqref{eq:SSym} and
  the matrix difference equations~\eqref{eq:QPeriod} if analogous
  relations hold for~$\tilde{f}^{1\cdots m}$. It was shown that
  solutions to the dimensional reduced
  problem can explicitly be constructed by use of the Bethe
  ansatz~\eqref{eq:BA.2}.

% -------------------------------------------------------------------------
% -------------------------------------------------------------------------
% -------------------------------------------------------------------------
\section{Highest-weight property}
  \label{ch:GroupTheory}

  We now investigate the $sl(2,1)$-properties of the shift operator
  $Q^{1\cdots n}$ and of the solutions~\eqref{eq:BA.1} constructed above.
  The behaviour $R_{ab}(x)=\Sigma_{ab}+\frac{K}{x}P_{ab}+O(x^{-2})$
  for $x\rightarrow\infty$ implies the asymptotic expansion
\begin{eqnarray*}
  [T_{1\cdots n,a}(\vec{x}|u)]_{\alpha,\{\gamma\}}^{\beta,\{\nu\}}
  &=&[\Sigma_{1a}\cdots\Sigma_{na}
     +\frac{K}{u}\sum_{j=1}^n\Sigma_{1a}\cdots P_{ja}\cdots\Sigma_{na}]_{\alpha,\{\gamma\}}^{\beta,\{\nu\}}
     +O(u^{-2})\\
  &=&\sigma_{\alpha,\{\mu\}}\delta_\alpha^\beta\delta_{\mu_1}^{\nu_1}\cdots\delta_{\mu_n}^{\nu_n}
     +\frac{K}{u}\sigma_{\beta\alpha}\sigma_{\beta,\{\nu\}}
     M_{\alpha,\{\mu\}}^{\beta,\{\nu\}}
     +O(u^{-2}).
\end{eqnarray*}
  The operators $M_{\alpha,\{\mu\}}^{\beta,\{\nu\}}$ have the form
\begin{eqnarray}
  \label{eq:GeneratorFormel}
  M_{\alpha,\{\mu\}}^{\beta,\{\nu\}}=\sum_j\sigma_{\beta\nu_{j+1}}\cdots\sigma_{\beta\nu_{n}}\sigma_{\alpha\nu_{j+1}}\cdots\sigma_{\alpha\nu_{n}}\delta_{\mu_1}^{\nu_1}\cdots\delta_{\mu_{j-1}}^{\nu_{j-1}}\delta_{\mu_j}^{\beta}\delta_{\alpha}^{\nu_j}\delta_{\mu_{j+1}}^{\nu_{j+1}}\cdots\delta_{\mu_n}^{\nu_n}.
\end{eqnarray}

  From this one derives the commutation relations
\begin{eqnarray}
  \label{eq:VTRMT}
   M_{\alpha}^{\alpha^\prime}\TS_{\beta}^{\beta^\prime}\!(u)
   -\sigma_{\alpha\beta}\sigma_{\alpha\beta^\prime}\sigma_{\alpha^\prime\beta}
    \sigma_{\alpha^\prime\beta^\prime}\TS_{\beta}^{\beta^\prime}\!(u) M_{\alpha}^{\alpha^\prime}
  =\delta_{\beta}^{\alpha^\prime}\TS_{\alpha}^{\beta^\prime}\!(u)
   -\sigma_{\alpha\beta}\sigma_{\alpha\beta^\prime}\sigma_{\alpha^\prime\beta}
    \sigma_{\alpha^\prime\beta^\prime}\delta_{\alpha}^{\beta^\prime}\TS_{\beta}^{\alpha^\prime}\!(u),
\end{eqnarray}
  where the quantum space indices have been neglected. A further consequence is
\begin{eqnarray*}
  M_{\alpha}^{\alpha^\prime}M_{\beta}^{\beta^\prime}
   -\sigma_{\alpha\beta}\sigma_{\alpha\beta^\prime}\sigma_{\alpha^\prime\beta}
    \sigma_{\alpha^\prime\beta^\prime}M_{\beta}^{\beta^\prime} M_{\alpha}^{\alpha^\prime}
  =\delta_{\beta}^{\alpha^\prime}M_{\alpha}^{\beta^\prime}
   -\sigma_{\alpha\beta}\sigma_{\alpha\beta^\prime}\sigma_{\alpha^\prime\beta}
    \sigma_{\alpha^\prime\beta^\prime}\delta_{\alpha}^{\beta^\prime}M_{\beta}^{\alpha^\prime}.
\end{eqnarray*}
  for~$u\rightarrow\infty$. This implies that the operators $M_{\alpha}^{\alpha^\prime}$
  are generators of~$sl(2,1)$ in the Cartan-Weyl basis
  (see~\cite{Foerster:1993uk,Scheunert:1977wj}).
  From~\eqref{eq:VTRMT} one can derive the invariance property
  $[M_{\alpha}^{\alpha^\prime},Q(\vec{u}|i)]_-=0$. This means that from any
  solution of~\eqref{eq:QPeriod} further solutions may be constructed by applying raising and
  lowering operators of~$sl(2,1)$. The operators
  $W_{\alpha}=M_{\alpha}^{\alpha}$ (no summation with respect to~$\alpha$)
  satisfy the commutation relations $[W_{\alpha},W_{\beta}]_{-}=0$ and
  generate the Cartan subalgebra. For~$\alpha=\beta$ the statistic signs 
  in~\eqref{eq:GeneratorFormel} cancel and therefore we get
\begin{eqnarray}
  %\label{eq:GewichtsOperator}
  [W_{\alpha}]_{\{\mu\}}^{\{\nu\}}=\sum_j\delta_{\mu_1}^{\nu_1}\cdots\delta_{\mu_{j-1}}^{\nu_{j-1}}\delta_{\mu_j}^{\alpha}\delta_{\alpha}^{\nu_j}\delta_{\mu_{j+1}}^{\nu_{j+1}}\cdots\delta_{\mu_n}^{\nu_n}.  
\end{eqnarray}

  The highest-weight property of the Bethe ansatz functions
  $M_{\alpha}^{\alpha^\prime}f^{1\cdots n}(\vec{x})=0$ for
  $\alpha^\prime>\alpha$ is proven in a way
  analogous to the one used in section~\ref{ch:MaDiffEq}. In other words one uses
  commutation relations implied by~\eqref{eq:VTRMT}, then commutes the
  matrices~$M_{\alpha}^{\alpha^\prime}$ through all $B$-operators to the right
  and finally, one uses certain eigenvalue equations. Again one has `wanted' and
  `unwanted' contributions and the summation guarantees the vanishing of the latter
  (compare~\cite{KarowskiSU(N)}). After some calculation one obtains the
  weight vector which is defined by $W_{\alpha}f(\vec{x})=w_{\alpha}f(\vec{x})$ and reads 
\begin{eqnarray*}
  \vec{w}=(n-m,m-k,k).
\end{eqnarray*}
  The highest-weight conditions are~$w_1\geq w_2\geq-w_3$
  and~$w_1,w_2,w_3\geq0$~\cite{Foerster:1993uk}.

% -------------------------------------------------------------------------
% -------------------------------------------------------------------------
% -------------------------------------------------------------------------
\section{Conclusions and outlook}
  \label{ch:Conclusions}

  In this article we have discussed a combined system of matrix difference
  equations based on the supersymmetric Lie algebra~$sl(2,1)$. Solutions 
  are constructed by means of a nested version of the so called off-shell
  Bethe ansatz and
  shown to be of highest weight with respect to~$sl(2,1)$. Due to the
  invariance of the shift operator~$Q^{1\cdots m}$ under
  the generators of~$sl(2,1)$ it is possible to construct and classify
  further solutions by purely group theoretic considerations.

  It would be interesting to see wether there is a quantum integrable
  (relativistic) field theory associated to the supersymmetric t-J model. 
  In that case the methods presented here could be used to determine the
  corresponding correlation functions. In this context the extension of our
  results to the $q$-deformed case~$U_q[sl(2,1)]$ would also be of
  interest~\cite{Foerster:1993fp,Foerster2}. Recently there has been discussed
  an integrable quantum field theory based on the $osp(2,2)$ graded Lie
  algebra~\cite{Bassi:1999ua} which is isomorphic to $sl(2,1)$.

% -------------------------------------------------------------------------
% -------------------------------------------------------------------------
% -------------------------------------------------------------------------
  \vspace{1cm}
{\!\!\!\bf Acknowledgements:}
  The author would like to thank M. Karowski, R. Schrader and in particular
  A. Zapletal for numerous helpful and stimulating discussions as well as
  Z. Maassarani for pointing out references~\cite{Bassi:1999ua, Ramos:1996my,
  Maassarani:1995ac} to him. 
  He is also grateful to Studienstiftung des deutschen Volkes for support.
  The work was partially supported by DFG, Sonderforschungsbereich 288
  `Differentialgeometrie und Quantenphysik'.

\bibliographystyle{utphys}
\bibliography{bibliography}

\end{document}